\documentclass[showpacs,preprintnumbers,amsmath,amssymb,floatfix]{revtex4}
\usepackage{graphicx}
\newcommand{\eq}[1]{Eq. (\ref{#1})}
\begin{document}

\title{Can gravitational dynamics be obtained by diffeomorphism invariance of action?}
\date{\today}
   \author{Sijie Gao}
    \email{sijie@bnu.edu.cn}
    \affiliation{Department of Physics, Beijing Normal University, Beijing, 100875,
    China}
    \author{Hongbao Zhang}
    \email{hbzhang@pkuaa.edu.cn}
    \affiliation{Department of Astronomy, Beijing Normal University, Beijing,
100875, China\\
Department of Physics, Beijing Normal University, Beijing, 100875,
China\\
CCAST (World Laboratory), P.O. Box 8730, Beijing, 100080, China}

\begin{abstract}
It has recently been suggested that the gravitational dynamics could
be obtained by requiring the action to be invariant under
diffeomorphism transformations. We argue that the action constructed
in usual way is automatically diffeomorphism invariant in nature,
which thus invalidates this alternative perspective to obtain
gravitational dynamics. Especially, we also show what is wrong with
the technical derivation of gravitational dynamics in the
alternative approach.
\end{abstract}


\maketitle As is well known, like any other classical field
equation, Einstein's equation can also be established by the
powerful variational principle. The corresponding action for gravity
is the so called Einstein-Hilbert one plus the ever neglected
surface term, i.e.,
\begin{equation}
S[g^{ab}]=\int_U R+2\int_{\partial U}K,\label{action}
\end{equation}
where $R$ is the Ricci scalar and $K$ is the trace of the extrinsic
curvature of the boundary\cite{Wald}.

Recently, Padmanabhan suggested that Einstein's equation plus an
undetermined cosmological constant can be derived from a new
perspective\cite{Pad, padgrqc}. Later Sotiriou and Liberati
reformulated this approach in \cite{Sot}. Here is the outline of the
reformulation in \cite{Sot}. Let $f_\lambda: U\rightarrow U$ be a
one-parameter family of diffeomorphisms. Imposing the diffeomorphism
invariance on this action, one has $S[g^{ab}]=S[(f_{\lambda*}
g)^{ab}]$, where $f_{\lambda*}$ is the usual push-forward mapping
associated with $f_\lambda$. Hence, for such variations, one obtains
\begin{equation}
0=\frac{dS}{d\lambda}=\int_U\frac{\delta S}{\delta
g^{ab}}\frac{dg^{ab}}{d\lambda}=\int_U\frac{\delta S}{\delta
g^{ab}}\mathcal{L}_\xi g^{ab}=2\int_U\frac{\delta S}{\delta
g^{ab}}\nabla^a\xi^b,
\end{equation}
where $\xi^b$ is the vector field, which generates $f_\lambda$, to
be parallel to the boundary. Later, if one further requires
$\mathcal{L}_\xi g^{ab}|_{\partial U}=0$,
i.e.,$\nabla_a\xi_b+\nabla_b\xi_a|_{\partial U}=0$, then
$\frac{\delta S}{\delta g^{ab}}$ reduces to $G_{ab}$, which follows
\begin{equation}
0=\int_UG_{ab}\nabla^a\xi^b=\int_U\nabla^a(G_{ab}\xi^b)-\int_U
(\nabla^aG_{ab})\xi^b=\int_{\partial U}G_{ab}n^a\xi^b-\int_U
(\nabla^aG_{ab})\xi^b, \label{several}
\end{equation}
where $n_a$ is the normal vector field to the boundary. Now taking
into account the fact that the above equation holds for arbitrary
$U$ and arbitrary vector field $\xi^a$ except that it satisfies both
$\xi^an_a=0$ and $\nabla_a\xi_b+\nabla_b\xi_a=0$ on the boundary,
the Bianchi identity can be obtained, i.e.,
\begin{equation}
\nabla^aG_{ab}=0,\label{equation1}
\end{equation}
which implies
\begin{equation}
\int_{\partial U}G_{ab}\xi^a n^b=0 \label{correct}.
\end{equation}
Especially, invoking $\xi^a$ arbitrary, apart from being equal to
the generator of so called Rindler horizon on the boundary, the
authors of \cite{Sot} claim that \eq{correct} leads to
\begin{equation}
G_{ab}=F(g)g_{ab},\label{equation2}
\end{equation}
where $F(g)$ is some scalar depending on the metric. Furthermore,
Using \eq{equation1}, and taking the covariant derivative of both
sides of \eq{equation2}, one obtains $F(g)=\Lambda$, an undetermined
cosmological constant. That is the sketch for the crucial point in
\cite{Sot}.

However, we stress that the action (\ref{action}) is manifestly
independent of choice of coordinate system, which is the alternative
representation for the automatical diffeomorphism invariance of the
action in the passive viewpoint\cite{Wald}. Therefore, the action
constructed from any Lorentzian metric is automatically invariant
under those diffeomorphisms mentioned before. Obviously, this
Lorentzian metric has not to satisfy Einstein's equation plus an
undetermined cosmological constant, which invalidates the
alternative derivation of Einstein's equation by diffeomorphism
invariance of the action\cite{footnote}. Virtually, not only the
action for gravity but also the action for matter fields is
automatically invariant under such diffeomorphisms, which
essentially originates from the fact that the involved fields are
all formulated in terms of tensor fields on a manifold\cite{Wald}.

Now we would also like to identify the loophole in the derivation of
\cite{Sot}. Note that \eq{correct}, which is essentially Eq. (17) in
\cite{Sot}, is correct. The key question is whether any meaningful
information can be extracted from it. The authors of \cite{Sot}
argued that \eq{correct} implies \eq{equation2} that leads to the
desired field equation. However, we shall show that \eq{correct}
gives no information on the metric. Let $\epsilon_{abcd}$ and
$\tilde \epsilon_{abc}$ be the volume elements on $U$ and $\partial
U$, respectively, such that the relation
\begin{equation}
\frac{1}{4} \epsilon_{abcd}=n_{[a}\tilde\epsilon_{bcd]}
\label{twoep}
\end{equation}
holds\cite{Wald}. Let $v^a=G^{ab}\xi_b$. Contracting $v^a$ into both
sides of \eq{twoep} and restricting the resulting 3-forms to vectors
only tangent to $\partial U$, one finds
\begin{equation}
\epsilon_{abcd}v^a=\left(n_av^a\right) \tilde\epsilon _{bcd}\,.
\label{contract}
\end{equation}
Note that the right-hand side of \eq{contract} is just the integrand
in \eq{correct}. Let $d$ be the derivative operator defined in
\cite{Wald}, mapping a $p-1$-form to a $p$-form. It is not difficult
to show (see also \cite{Wald}) that
\begin{equation}
d_e(\epsilon_{abcd}v^a)=(\nabla_a v^a)\epsilon_{ebcd}\,.
\end{equation}
Using the Bianchi identity and the Killing equation satisfied by
$\xi^a$, we have
\begin{equation}
\nabla_a v^a=\nabla_a(G^{ab}\xi_b)=0\,.
\end{equation}
This shows that the left-hand side of \eq{contract} is a closed
form. According to the results in \cite{jmp}(see also \cite{Wald}),
such a form must be exact, i.e., there exists a form $\beta_{cd}$
such that
\begin{equation}
\epsilon_{abcd}v^a=d_b\beta_{cd}\,.
\end{equation}
Thus, the integrand of left-hand side of \eq{correct} is an exact
form. Consequently, the integral vanishes if $\partial U$ is a
compact boundaryless surface (this is the case relevant to our
discussion). Therefore, \eq{correct} always holds for any metric for
which the Killing equation is satified on the boundary, and such
nontrivial information as \eq{equation2} on the metric can not be
provided by \eq{correct}.

In conclusion, this paper demonstrates that the alternative
derivation of gravitational dynamics by diffeomorphism invariance of
the action is infeasible. We accomplish it by both making it obvious
that the diffeomorphism invariance of the covariant action holds
automatically for any metric and pointing out the mistake made in
the alternative derivation.
\section*{Acknowledgements}
We would like to thank Professor C. Liang for helpful discussions.
S. Gao was supported in part by NSFC(No.10605006). H. Zhang was
supported in part by NSFC(No.10533010).


\begin{thebibliography}{0}
\bibitem{Wald}R. M. Wald, General Relativity(The University of Chicago Press, Chicago, 1984).
\bibitem{Pad}T. Padmanabhan, Int. J. Mod. Phys. D14 2263(2005).
\bibitem{padgrqc} T. Padmanabhan, gr-qc/0609012.
\bibitem{Sot}T. P. Sotiriou and S. Liberati, Phys. Rev. D74
044016(2006).
\bibitem{footnote}Although the arguments and
results in \cite{Sot} appear to be similar to those in \cite{Pad} or
\cite{padgrqc}, our criticisms in this paper do not apply to the
latter due to some essential distinctions in their derivations. 
\bibitem{jmp} R. M. Wald, J. Math. Phys. 31 2378(1990).
\end{thebibliography}
\end{document}